\documentclass[10pt,journal,compsoc]{IEEEtran}
\usepackage{algorithm}
\usepackage{graphicx}
\usepackage[cmex10]{amsmath}
\usepackage{algorithmic}
\usepackage{amssymb}
\usepackage{amsfonts}
\usepackage{dsfont}
\usepackage{array}
\usepackage{fixltx2e}
\usepackage{dblfloatfix}
\usepackage{color}
\usepackage{paralist,xparse}
\usepackage{adjustbox,lipsum}
\usepackage{mathtools}
\usepackage{amsmath,epsfig,subfigure,paralist,verbatim,tikz}
\usepackage[hyphens]{url}
\usepackage[european resistor, european voltage, european current]{circuitikz}
\usetikzlibrary{matrix,arrows,shapes,positioning,snakes}
\usetikzlibrary{decorations.markings,decorations.pathmorphing,decorations.pathreplacing}
\usetikzlibrary{calc,patterns,shapes.geometric,topaths,fit}
\usepackage{amsfonts,nth}
\usepackage{epsfig,psfrag}
\usepackage[font=small,labelfont=bf]{caption}
\usepackage{epstopdf}
\usepackage{pgfplots}
\usepgfplotslibrary{dateplot}
\usepackage{multirow}
\usepackage{rotating}
\newlength\figureheight
\newlength\figurewidth
\usepackage{tabu}
\usepackage{booktabs}
\usepackage{amsmath,amssymb}
\usepackage{dsfont}
\usepackage{cancel}
\usepackage{graphicx}
\usepackage{xargs}
\usepackage{xspace}

\def\qed{\ifhmode\unskip\nobreak\fi\hfill \ensuremath{\square}}

\newcommand{\Exp}[1]{\ensuremath{\exp\left\{#1\right\}}}

\newcommand{\hyp}[2]{
\ensuremath{H_0:#1 \ifhmode\quad\text{versus}\quad\fi\text{ vs. } H_1:#2}}

\newcommandx{\unif}[1][1={a,b}]{\textrm{Unif}\left({#1}\right)}
\newcommandx{\unifd}[1][1={a,\ldots,b}]{\textrm{Unif}\left\{{#1}\right\}}
\newcommandx{\dunif}[3][1=x,2=a,3=b]{\frac{I(#2<#1<#3)}{#3-#2}}
\newcommandx{\dunifd}[3][1=x,2=a,3=b]{\frac{I(#2\le#1\le#3)}{#3-#2+1}}
\newcommandx{\punif}[3][1=x,2=a,3=b]{
\begin{cases} 0 & #1 < #2 \\ \frac{#1-#2}{#3-#2} & #2 < #1 < #3 \\ 1 & #1 > #3\\\end{cases}}
\newcommandx{\punifd}[3][1=x,2=a,3=b]{
\begin{cases} 0 & #1 < #2\\ \frac{\lfloor#1\rfloor-#2+1}{#3-#2} & #2 \le #1 \le #3 \\ 1 & #1 > #3\\ \end{cases}}

\newcommandx\bern[1][1=p]{\textrm{Bern}\left({#1}\right)}
\newcommandx\dbern[2][1=x,2=p]{#2^{#1} \left(1-#2\right)^{1-#1}}
\newcommandx\pbern[2][1=x,2=p]{\left(1-#2\right)^{1-#1}}

\newcommandx\bin[1][1={n,p}]{\textrm{Bin}\left(#1\right)}
\newcommandx\dbin[3][1=x,2=n,3=p]{\binom{#2}{#1}#3^#1\left(1-#3\right)^{#2-#1}}

\newcommandx\mult[1][1={n,p}]{\textrm{Mult}\left(#1\right)}
\newcommandx\dmult[3][1=x,2=n,3=p]{\frac{#2!}{#1_1!\ldots#1_k!}#3_1^{#1_1}\cdots#3_k^{#1_k}}

\newcommandx\hyper[1][1={N,m,n}]{\textrm{Hyp}\left({#1}\right)}
\newcommandx\dhyper[4][1=x,2=N,3=m,4=n]{\frac{\binom{#3}{#1}\binom{#2-#3}{#4-#1}}{\binom{#2}{#4}}}

\newcommandx\nbin[1][1={r,p}]{\textrm{NBin}\left({#1}\right)}
\newcommandx\dnbin[3][1=x,2=r,3=p]{\binom{#1+#2-1}{#2-1}#3^#2(1-#3)^#1}
\newcommandx\pnbin[3][1=x,2=r,3=p]{I_#3(#2,#1+1)}

\newcommandx\geo[1][1=p]{\textrm{Geo}\left(#1\right)}
\newcommandx\dgeo[2][1=x,2=p]{#2(1-#2)^{#1-1}}
\newcommandx\pgeo[2][1=x,2=p]{1-(1-#2)^#1}

\newcommandx\pois[1][1=\lambda]{\textrm{Po}\left({#1}\right)}
\newcommandx\dpois[2][1=x,2=\lambda]{\frac{#2^#1 e^{-#2}}{#1!}}
\newcommandx\ppois[2][1=x,2=\lambda]{e^{-#2}\sum_{i=0}^#1\frac{#2^i}{i!}}

\newcommandx\norm[1][1={\mu,\sigma^2}]{\mathcal{N}\left({#1}\right)}
\newcommandx\dnorm[3][1=x,2=\mu,3=\sigma]%
{\frac{1}{#3\sqrt{2\pi}}\Exp{-\frac{\left(#1-#2\right)^2}{2 #3^2}}}
\newcommandx\pnorm[1][1=x]{\Phi\left({#1}\right)}
\newcommandx\qnorm[1]{\Phi^{-1}\left({#1}\right)}

\newcommandx\mvn[1][1={\mu,\Sigma}]{\mathrm{MVN}\left({#1}\right)}

\newcommandx\ex[1][1=\beta]{\textrm{Exp}\left(#1\right)}
\newcommandx\dex[2][1=x,2=\beta]{\frac{1}{#2}e^{-#1/#2}}
\newcommandx\pex[2][1=x,2=\beta]{1-e^{-#1/#2}}

\newcommandx\gam[1][1={\alpha,\beta}]{\textrm{Gamma}\left({#1}\right)}
\newcommandx\dgamma[3][1=x,2=\alpha,3=\beta]%
{\frac{#3^{#2}}{\Gamma\left( #2 \right)} #1^{#2-1}e^{-#3#1}}

\newcommandx\invgamma[1][1={\alpha,\beta}]{\textrm{InvGamma}\left({#1}\right)}
\newcommandx\dinvgamma[3][1=x,2=\alpha,3=\beta]%
{\frac{#3^{#2}}{\Gamma\left(#2\right)}#1^{-#2-1}e^{-#3/#1}}
\newcommandx\pinvgamma[3][1=x,2=\alpha,3=\beta]%
{\frac{\Gamma\left(#2,\frac{#3}{#1}\right)}{\Gamma\left(#2\right)}}

\newcommandx\bet[1][1={\alpha,\beta}]{\textrm{Beta}\left(#1\right)}
\newcommandx\dbeta[3][1=x,2=\alpha,3=\beta]
{\frac{\Gamma\left(#2+#3\right)}{\Gamma\left(#2\right)\Gamma\left(#3\right)}#1^{#2-1}\left(1-#1\right)^{#3-1}}

\newcommandx\dir[1][1={\alpha}]{\textrm{Dir}\left(#1\right)}
\newcommandx\ddir[3][1=x,2=\alpha]{\frac{\Gamma\left(\sum_{i=1}^k #2_i\right)}{\prod_{i=1}^k\Gamma\left(#2_i\right)}\prod_{i=1}^k #1_i^{#2_i-1}}

\newcommandx\weibull[1][1={\alpha}]{\textrm{Dir}\left(#1\right)}
\newcommandx\dweibull[3][1=x,2=\lambda,3=k]{\frac{#3}{#2}
\left(\frac{#1}{#2}\right)^{#3-1} e^{-(#1/#2)^k}}

\newcommandx\chisq[1][1=k]{\chi_{#1}^2}

\newcommandx\zet[1][1=s]{\textrm{Zeta}\left(#1\right)}
\newcommandx\dzeta[2][1=x,2=s]{\frac{#1^{-#2}}{\zeta\left(#2\right)}}

\newcommandx\AR[1][1=p]{\mathsf{AR}\left({#1}\right)}
\newcommandx\MA[1][1=q]{\mathsf{MA}\left({#1}\right)}
\newcommandx\ARMA[1][1={p,q}]{\mathsf{ARMA}\left({#1}\right)}
\newcommandx\ARIMA[1][1={p,d,q}]{\mathsf{ARIMA}\left({#1}\right)}
\newcommandx\SARIMA[3][1={p,d,q},2={P,D,Q},3=s]{\mathsf{ARIMA}\left(#1\right) \times \left(#2\right)_{#3}}

\newcommandx\step[1][1=t]{^{(#1)}}

\newcommand{\metalevel}{\text{meta-level }}
\newcommand{\viewcount}{\text{view count }}
\newcommand{\viewcounts}{\text{view count }}
\usepackage{footmisc}
\graphicspath{{./figs/}}
\usepackage[hidelinks]{hyperref}
\usepackage{pstool}
\ifCLASSOPTIONcompsoc
  \usepackage[nocompress]{cite}
\else
  \usepackage{cite}
\fi
\newif\ifBBTV
\BBTVfalse
\ifBBTV
\usepackage[compact]{titlesec}
\usepackage{fullpage}

\else
\fi

\begin{document}
\title{Engagement dynamics and sensitivity analysis of YouTube videos}
\author{William~Hoiles,~\IEEEmembership{Student~Member,~IEEE,}
        Anup~Aprem,
        and~Vikram~Krishnamurthy,~\IEEEmembership{Fellow,~IEEE}
\IEEEcompsocitemizethanks{\IEEEcompsocthanksitem The authors are with the Department
of Electrical and Computer Engineering, University of British Columbia, Vancouver,
Canada.\protect\\
E-mail: $\{$whoiles,aaprem,vikramk$\}$\@ece.ubc.ca
}
}

\markboth{}{Engagement dynamics and sensitivity analysis of YouTube videos}
\IEEEtitleabstractindextext{%
\begin{abstract}
	YouTube, with millions of content creators, has become the preferred destination for watching videos online. Through the Partner program, YouTube allows content creators to monetize their popular videos. Of significant importance for content creators is which \metalevel features (e.g.\ title, tag, thumbnail) are most sensitive for promoting video popularity. The popularity of videos also depends on the social dynamics, i.e.\ the interaction of the content creators (or channels) with YouTube users. Using real-world data consisting of about $6$ million videos spread over $25$ thousand channels, we empirically examine the sensitivity of YouTube \metalevel features and social dynamics. The key meta-level features that impact the view counts of a video include: first day \viewcount, number of subscribers, contrast of the video thumbnail, Google hits, number of keywords, video category, title length, and number of upper-case letters in the title respectively and illustrate that these meta-level features can be used to estimate the popularity of a video. In addition, optimizing the \metalevel features after a video is posted increases the popularity of videos. In the context of social dynamics, we discover that there is a causal relationship between views to a channel and the associated number of subscribers. Additionally, insights into the effects of scheduling and video playthrough in a channel are also provided. Our findings provide a useful understanding of user engagement in YouTube. 
\end{abstract}

}

\maketitle

\IEEEdisplaynontitleabstractindextext
\IEEEpeerreviewmaketitle

\IEEEraisesectionheading{\section{Introduction}\label{sec:introduction}}
The YouTube social network contains over 1 billion users who collectively watch millions of hours of YouTube videos and generate billions of views every day. Additionally, users upload over $300$ hours of video content every minute. YouTube generates billions in revenue through advertising and through the Partner program shares the revenue with the content creators. 

The video view count is a key metric of the measure of popularity or ``user engagement'' of a video and the metric by which YouTube pays the content providers\footnote{However, recently, view time is gaining more prominence than view count.}. 
A key question is: {\em How do \metalevel features of a posted video (e.g. thumbnail, title, tags, description) drive user engagement in the YouTube social network? } 
However, the content alone does not influence the popularity of a video. 
YouTube also has a social network layer on top of it's media content. 
The main social component is how the content creators (also called ``channels'') interact with the users. 
So another key question is: {\em How does the interaction of the YouTube channel with the user affect popularity of videos? }
In this paper, we study both the above questions. In particular, our aim is to examine how the individual video features (through the \metalevel data) and the social dynamics contribute to the popularity of a video. 

\subsection*{Literature Review}
The study of popularity of YouTube videos based on \metalevel features is a challenging problem given the diversity of users and content providers. Several models on characterizing the popularity of YouTube videos are parametric in form, where the view count time series is used to estimate the model parameters. For example, ARMA time series models~\cite{GCM11}, multivariate linear regression models~\cite{PAG13}, modified Gompertz models~\cite{RAEJLP14,REJAL15}, have been utilized to estimate the future video view counts given past view count time series. Using only the title of the video (one of the \metalevel features) \cite{Zha15} considers the problem of predicting whether the view count will be high or low. In a related context, \cite{YSA14,YHAM15} studied the importance of tags for Flicker data. Aside from text based \metalevel features (title and tags), in~\cite{TR15} Support Vector Regression (SVR) is proposed to predict the popularity using features of the video frames (e.g. face present, rigidity, color, clutter). It is illustrated in~\cite{TR15} that using the combination of visual features and temporal dynamics results in improved performance of the SVR for predicting \viewcounts compared to using only visual features or temporal dynamics alone. In the social context, the uploading behaviour of YouTube content creators was studied in~\cite{Ding2011}. Specifically, the paper finds that YouTube users with a social network are more popular compared to other users. 

\subsection*{Main results} 
In this paper, we investigate how the \metalevel features and the interaction of the YouTube channel with the users affect the popularity of videos. For convenience we summarize the main empirical conclusions of this paper:
\begin{compactenum}
\item The five dominant meta-level features that affect the popularity of a video are: first day \viewcount, number of subscribers, contrast of the video thumbnail, Google hits, and number of keywords. Sec.~\ref{sec:sensitivity:analysis:elm} discusses this further.  
\item Optimizing the meta-level features (e.g.\ thumbnail, title, tags, description) after a video has been posted increases the popularity of the video. In addition, optimizing the title increases the traffic due to YouTube search, optimizing the thumbnail increases the traffic from related videos and optimizing the keywords increases the traffic from related and promoted videos. Sec.~\ref{subsec:sensitivity:metalevel:optimization} provides details on this analysis.
\item Insight into the causal relationship between the subscribers and view count for YouTube channels is also explored. For popular YouTube channels, we found that the channel view count affects the subscriber count, see Sec.~\ref{sec:causal:relation:YouTube}. 
\item New insights into the scheduling dynamics in YouTube gaming channels are also found. For channels with a dominant periodic uploading schedule, going ``off the schedule'' increases the popularity of the channel, see Sec.~\ref{sec:scheduling:YouTube}. 
\item The generalized Gompertz model can be used to distinguish views due to virality (views from subscribers), migration (views from non-subscribers) and exogenous events, see Sec.~\ref{sec:seperate:virality:migration}. 
\item New insights into playlist dynamics. The early view count dynamics of a YouTube videos are highly correlated with the long term ``migration'' of viewers to the video. Also, early videos in a game playthrough typically contain higher views compared with later videos in a game playthrough playlist, see Sec.~\ref{sec:playlist:dynamics}. 
\item The number of subscribers of a channel only affects the early view count dynamics of videos in a playthrough, see Sec.~\ref{sec:playlist:dynamics}. 
\end{compactenum} 
All the above results are validated on the YouTube dataset consisting of over $6$ million videos across $25$ thousand channels.  This dataset\footnote{The Appendix summarizes the key features of the YouTube dataset that we have used. } was provided to us by BroadbandTV Corp.\ (BBTV).  
The dataset consists of daily samples of metadata of the YouTube videos on the BBTV platform from April, 2007 to May, 2015. BBTV is one of the largest Multi-channel network (MCN) in the world\footnote{\url{http://variety.com/2016/digital/news/broadbandtv-mcn-disney-maker-comscore-1201696857/}}. 
The results of the paper allows YouTube partners such as BBTV to adapt their user engagement strategies to generate more views and hence increase revenue. 

{\bf Caveat}: It is important to note that the above empirical conclusions are based on the BBTV dataset. These videos cover the YouTube categories of gaming, entertainment, food, music, and sports as described in Table~\ref{tab:category:dataset:summary} of the Appendix. 
Whether the above conclusions hold for other types of YouTube videos is an open issue that is beyond the scope of this paper.  


The organization of the paper is as follows. 
In Sec.~\ref{sec:sensitivity:analysis:elm}, we use several machine learning methods to characterize the sensitivity of meta-level features on the popularity of YouTube videos. 
In Sec.~\ref{sec:social:dynamics:YouTube}, we use time series analysis methods to investigate how the interaction of content creators with users affect the popularity of videos. 
Using Granger causality, we determine the causal relation between \viewcounts and subscribers for channels in Sec.~\ref{sec:causal:relation:YouTube}.  
Sec.~\ref{sec:scheduling:YouTube} studies the scheduling dynamics of YouTube channels. 
Sec.~\ref{sec:seperate:virality:migration} addresses the problem of separating the \viewcount dynamics due to virality (\viewcount resulting from subscribers), migration (views from non-subscribers) and exogenous events (events other than \metalevel optimization considered in Sec.~\ref{sec:sensitivity:analysis:elm}), which affect the popularity of the videos using a generalized Gompertz model. 
In Sec.~\ref{sec:playlist:dynamics}, we study the playlist dynamics in YouTube. 

\section{Sensitivity Analysis of YouTube Meta-Level Features}
\label{sec:sensitivity:analysis:elm}
\begin{figure}[h]
	\centering
	\includegraphics[angle=0,width=1.5\figurewidth]{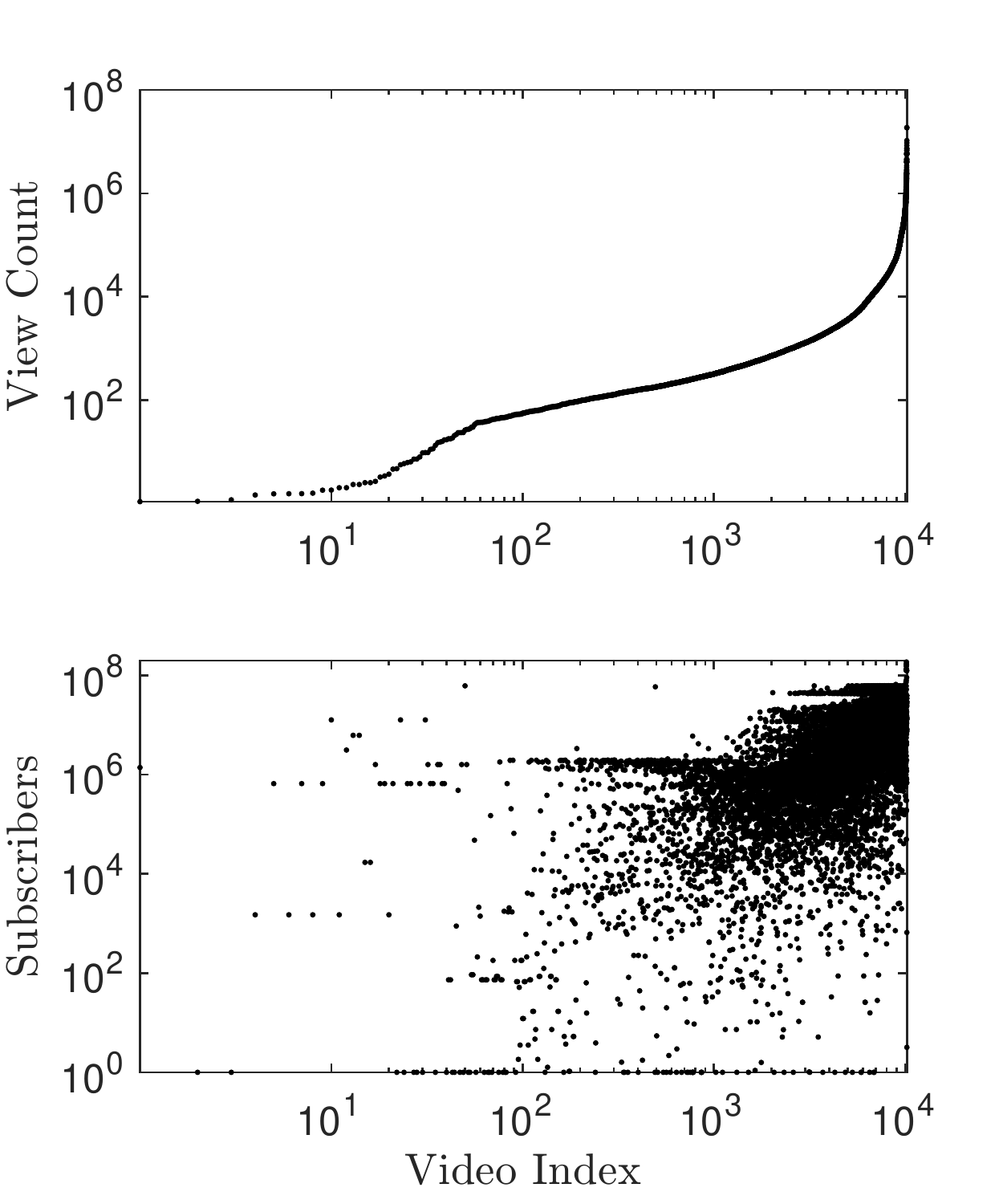}
	\caption{The top figure shows the view count of all videos (arranged according to increasing order of view count) after 14 days of the video being posted. The bottom figure shows the associated subscriber count when the video was posted.}
	\label{fig:subscribersandviewcount}
\end{figure}

In this section we apply machine learning methods to study how meta-level features of YouTube videos impacts the \viewcount of the video. The machine learning methods we utilize include: the Extreme Learning Machine (ELM)~\cite{HSZS13,BSZLH13}, Feed-Forward Neural Network~\cite{Bis06}, Stacked Auto-Encoder Deep Neural-Network~\cite{HZZW15,Sch15}, Elasticnet~\cite{ZH05}, Lasso, Relaxed Lasso~\cite{Mei07}, Quantile Regression with Lasso~\cite{PW15}, Conditional Inference Random Forest~\cite{HHZ06}, Boosted Generalized Additive Model~\cite{VR13,HBKSH10}, Bagged MARS using gCV Pruning~\cite{Dru97}, Generalized Linear Model with Stepwise Feature Selection using Akaike information criterion, and Spike and Slab Regression~\cite{IR05}. Additionally we use the feature selection method Hilbert-Schmidt Independence Criterion Lasso (HSIC-Lasso)~\cite{YJSXS14} to study the sensitivity of meta-level features which may be highly correlated. Note that the meta-level features used for prediction in the YouTube dataset contain significant noise. For example,  Fig.~\ref{fig:subscribersandviewcount} illustrates a trace of the subscribers when the video was posted, and the associated viewcount 14 days after the video has been posted. Therefore the machine learning algorithms used must be able to address this challenging problem of mapping from these type of noisy meta-level features to the associated \viewcount of a video. Of these methods we found that the ELM provides sufficient performance to both be used to estimate the meta-level features which significantly contribute to the \viewcount of a video, and for predicting the \viewcount of videos. 

\subsection{Extreme Learning Machine}
The dataset of features (described in Sec.~\ref{subsec:ELMsensitivity}) and \viewcounts are denoted as $\mathcal{D}=\{(x_i,v_i)\}_{i=1}^N$ where $x_i\in\mathds{R}^m$ is the feature vector, of dimension $m$, for video $i$, and $v_i$ is the total \viewcount for video $i$. Here, $N$ is the number of videos in the training dataset (The ELM was trained for three categories of videos, for details see Sec.~\ref{subsec:ELMsensitivity}). The ELM is a single hidden-layer feed-forward neural network--that is, the ELM consists of an input layer, a single hidden layer of $L$ neurons and an output layer. 
Each hidden-layer neuron can have a unique transfer function. Popular transfer functions include the sigmoid, hyperbolic tangent, and Gaussian. However any non-linear piecewise continuous function can be utilized. The output layer is obtained by a weighted linear combination of the output of the $L$ hidden neurons. 

The ELM model presented in~\cite{HZS06,HC07} is given by:
\begin{equation}
v_i=\sum_{k=1}^L \beta_kh_k(x_i;\theta_k),
\label{eqn:elm}
\end{equation}
where $\beta_k$ is the weight of neuron $k$, and $h_k(\cdot;\theta_k)$ is the hidden-layer neuron transfer function with parameter $\theta_k$, and $L$ is the total number of hidden-layer neurons in the ELM. Given $\mathcal{D}$, how can the ELM model parameters $\beta_k,\theta_k,$ and $L$ in (\ref{eqn:elm}) be selected? Given $L$, the ELM trains $\beta_k$ and $\theta_k$ in two steps. First, the hidden layer parameters $\theta_k$ are randomly initialized. Any continuous probability distribution can be used to initialize the parameters $\theta_k$. Second, the parameters $\beta_k$ are selected to minimize the square error between the model output and the measured output from $\mathcal{D}$. Formally, 
\begin{equation}
\beta^*\in\underset{\beta\in\mathds{R}^L}{\operatorname{arg max}}\Big\{||H\beta-V||^2_2\Big\},
\label{eqn:elmtrain}
\end{equation}
where $H$ denotes the hidden-layer output matrix with entries $H_{kj}=h_k(x_j;\theta_k)$ for $k\in\{1,2,\dots,L\}$ and $j\in\{1,2\dots,N\}$, and $V$ the target output with entries $V=[v_1, v_2,\dots,v_N]$. The solution to (\ref{eqn:elmtrain}) is given by $\beta^*=H^\dagger V$ where $H^\dagger$ denotes the Moore-Penrose generalized inverse of $H$. The major benefit of using the ELM, compared to other single layer feed-forward neural network, is that the training only requires the random generation of the parameters $\theta_k$, and the parameters $\beta_k$ can be computed as the solution of a set of linear equations. The computational cost of training the ELM is $O(N^3)$ for constructing the Moore-Penrose inverse~\cite{GV12}. 

\subsection{Sensitivity Analysis (Background)}
\label{subsec:sensitivity:analysis:background}
There are several sensitivity analysis techniques available in the literature~\cite{LM12,SJ15} which can be classified into two groups: filter methods, and wrapper methods. 
The filter methods consider only the meta-level features and the viewcount without the information available from a machine learning algorithm. The wrapper methods, on the other hand, utilize the information from the machine learning algorithm. Typically, wrapper methods give a more accurate measure of the sensitivity compared to filter methods~\cite{LM12,SJ15}.  
However, filter methods are computationally less expensive than wrapper methods and do not require the training and evaluation of the machine learning algorithm. Given the noise present in the meta-level features (Fig.~\ref{fig:subscribersandviewcount}) and the non-linearity between the meta-level features and \viewcount, filter methods are not suitable for the sensitivity analysis of the meta-level features. Hence, in this section we focus on two wrapper methods suitable for estimating the sensitivity of meta-level features on the \viewcount of YouTube videos.

For the first method we focus on the ELM (\ref{eqn:elm}) for evaluating the sensitivity of the meta-level features, however the method can be used for any machine learning method. Given that the ELM (\ref{eqn:elm}) is a single feed-forward hidden layer neural network, it is possible to evaluate the sensitivity of the meta-level features by taking the partial derivative of (\ref{eqn:elm}) for the trained ELM. Note that this method is utilized to estimate the sensitivity of input features in neural networks~\cite{GDL03}. The \emph{sum of squares derivatives}, denoted by $\text{SSD}_k$ for meta-level feature $x(k)$, is given by:
\begin{align}
\text{SSD}_k &= \sum_{i=1}^N\Big(\frac{\partial v_i}{\partial x(k)}\Big)^2 = \sum_{i=1}^N\Big(\sum_{k=1}^L \beta_k\frac{\partial h_k(x_i;\theta_k)}{\partial x(k)}\Big)^2.
\label{eqn:padsensitivity}
\end{align}
The variable with the largest $\text{SSD}_k$ is most influential to the prediction of the \viewcount using the ELM $v$ (\ref{eqn:elm}). Note that since the ELM is trained using all the meta-level features, the $\text{SSD}_k$ evaluates the average sensitivity of changes in a single meta-level feature with all other features held constant. 

To account for significant interdependency relationships between meta-level features requires sophisticated methods to evaluate the meta-level feature sensitivities. A state-of-the art method which can be used for this task is the Hilbert-Schmidt Independence Criterion Lasso (HSIC-Lasso)~\cite{YJSXS14}. The main idea of this method is to use the benefits of least absolute shrinkage and selection operator (Lasso) with a feature wise kernel to capture the non-linear input-output dependency. A measure of the importance of a meta-level feature is then given by the coefficient of the centered Gram matrix used in the HSIC-Lasso. 

Both of these methods will be applied to the YouTube dataset to study the sensitivity of the meta-level features of YouTube videos on the videos \viewcount. 

\subsection{Sensitivity of YouTube Meta-Level Features and Predicting View Count}

\label{subsec:ELMsensitivity}
In this section, the ELM (\ref{eqn:elm}) and other state-of-the art machine learning methods are applied to the YouTube dataset to compute the sensitivity of a videos meta-level features on the \viewcount of the video based on the feature importance measure $\text{SSD}_k$ (\ref{eqn:padsensitivity}). Videos of different popularity, (i.e.\ highly popular, popular, and unpopular as defined in Table~\ref{tab:popularity:dataset:summary} in the Appendix), may have different sensitivities to the meta-level features. Hence, in this paper, we independently perform the sensitivity analysis on the three popularity categories. First we define the meta-level features for each video, then evaluate the meta-level feature sensitivities on the associated \viewcount, and finally provide methods to predict the \viewcount of YouTube videos using various machine learning techniques. The analysis provides insight into which meta-level features are useful for optimizing the \viewcount of a YouTube video.

\subsubsection{Meta-Level Feature Construction}
\label{subsubsec:metalevelfeaturedataset}
Each YouTube video contains four primary components: the Thumbnail of the video, the Title of the video, the Keywords (also known as tags), and the description of the video. However, in typical user searches only a subset of the description is provided to the user. Therefore, we do not consider the contents of the description to significantly affect the \viewcount of the video. The meta-level features are constructed using the Thumbnail, Title, and Keywords. For the Thumbnail, 19 meta-level features are computed which include: the blurriness (e.g. CannyEdge, Laplace Frequency), brightness, contrast (e.g. tone), overexposure, and entropy of the thumbnail. For the Title, 23 meta-level features are computed which include: word count, punctuation count, character count, Google hits (e.g. if the title is entered into the Google search engine how many results are found), and the Sentiment/Subjectivity of the title computed using Vader~\cite{HG14}, and TextBlob~\footnote{\url{http://textblob.readthedocs.io/en/dev/}}. For the Keywords, 7 meta-level features are computed which include: the number of keywords, and keyword length. In addition, to the above 49 meta-level features, we also include auxiliary user meta-level features including: the number of subscribers,  resolution of the thumbnail used, category of the video, the length of the video, and the first day \viewcount of the video. Note that our analysis does not consider the video or audio quality of the YouTube video. 
Our analysis is focused on the sensitivity of the \viewcount based on the Thumbnail, Title, Keywords, and auxiliary channel information of the user that uploaded the video. In total $54$ meta-level features are computed for each video. The complete dataset used for the sensitivity analysis is given by $\mathcal{D}=\{(x_i,v_i)\}_{i=1}^N$, with $x_i\in\mathds{R}^{54}$ the computed meta-level features for video $i\in\{1,\dots,N\}$, $v_i$ the \viewcount 14 days after the video is published, and $N=10^4$, the total number of videos used for the sensitivity analysis. Note that the \viewcount $v_i$ is on the log scale (i.e. if a video has $10^6$ views then $v_i=6$). This is a necessary step as the range of view counts is from $10^2$ to above $10^7$. 

Prior to performing any analysis we pre-process the meta-level features in the dataset $\mathcal{D}$. First, all the meta-level features are scaled to satisfy $x(k)\in[0,1]$. Note that the meta-level features were not whitened (e.g. the meta-level data as not transformed to have an identity covariance matrix). The second pre-processing step involves removing redundant features in $\mathcal{D}$. Feature selection is a popular method for eliminating redundant meta-level features. In this paper we employ a correlation based feature selection based on the Pearson correlation coefficient (which was used for feature selection in~\cite{Hall99}) to eliminate the redundant meta-level features. Of the original 54 meta-level features, $m=29$ meta-level features remain after the removal of the correlated meta-level features. Note that removal of these features does not significantly impact the performance of the machine learning algorithms or the sensitivity analysis results.

\subsubsection{Meta-Level Feature Sensitivity}
\begin{figure}[h!]
\centering
\includegraphics[angle=0,clip=1,trim=60 0 70 0,width=3.2in]{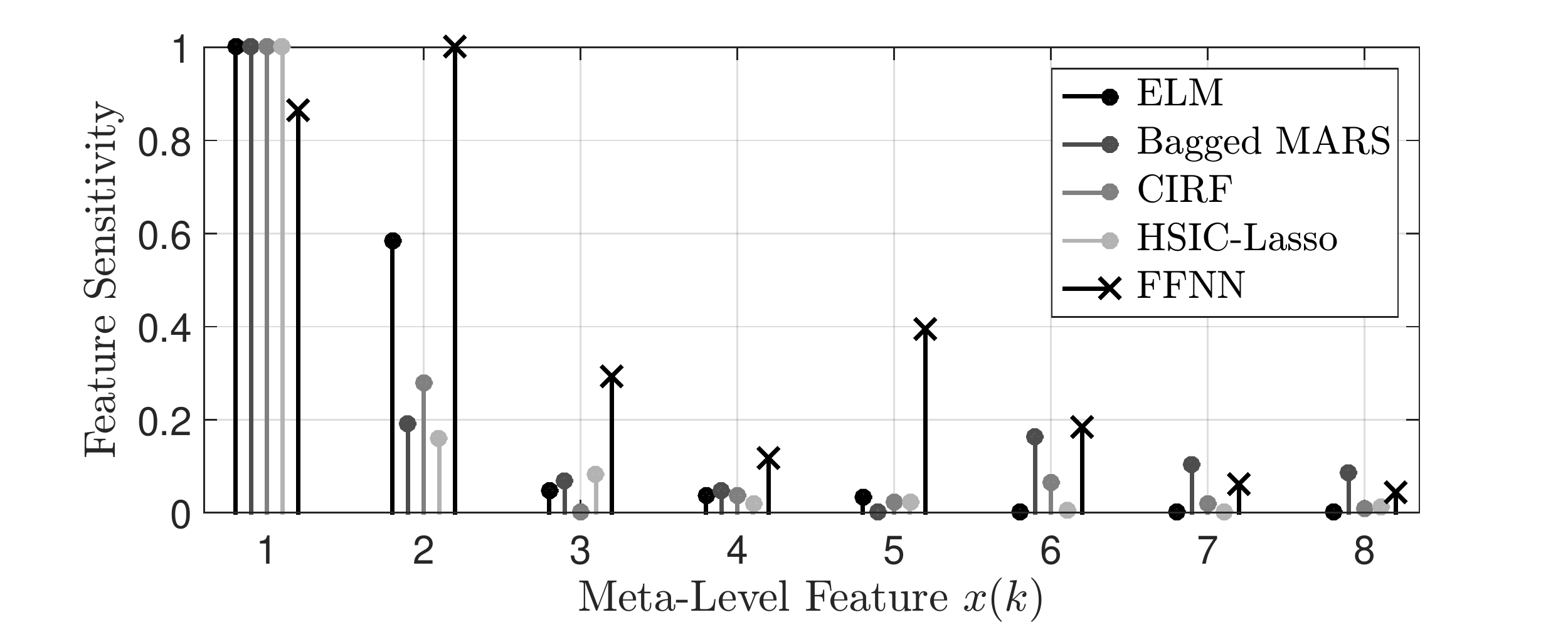}
\caption{Sensitivity of the meta-level features computed using the sum of squares derivatives $\text{SSD}_k$ (\ref{eqn:padsensitivity}) for the ELM, Bagged MARS, CIRF, and FFN, and the associated  coefficient of the centered Gram matrix for the HSIC-Lasso using the dataset $\mathcal{D}$ defined in Sec.\ref{subsubsec:metalevelfeaturedataset}. The meta-level features k=1 to k=8 are associated with: first day \viewcount, number of subscribers, contrast of the video thumbnail, Google hits, number of keywords, video category, title length, and number of upper-case letters in the title respectively. Similar results are obtained for highly popular, popular, and unpopular videos as defined in Table~\ref{tab:popularity:dataset:summary}.}
\label{fig:ELMperformancesensitivity}
\end{figure}

Given the dataset $\mathcal{D}=\{(x_i,v_i)\}_{i=1}^N$ constructed in Sec.\ref{subsubsec:metalevelfeaturedataset}, the goal is to estimate which features significantly contribute to the \viewcount of a video. To perform this sensitivity analysis five machine learning algorithms which include: the ELM, Bagged MARS using gCV Pruning~\cite{Dru97}, Conditional Inference Random Forest (CIRF)~\cite{HHZ06}, Feed-Forward Neural Network (FFNN)~\cite{Bis06}, and the feature selection method Hilbert-Schmidt Independence Criterion Lasso (HSIC-Lasso)~\cite{YJSXS14}. Each of these models is trained using a 10-fold cross validation technique, and the design parameters of each was optimized via extensive empirical evaluation. We selected the ELM (\ref{eqn:elm}) to contain $L=100$ neurons which ensures that we have sufficient accuracy on the predicted \viewcount given the features $x_i$, while reducing the effects of over-fitting. For the CIRF the design parameter for randomly selected predictors was set to 6, and the FFNN we have 10 neurons in the hidden-layer. The HSIC-Lasso regularization parameter was set to 100. Given the trained models, the sensitivity of the \viewcount on the meta-level features of a video is computed by evaluating the sum of squares derivatives, $\text{SSD}_k$ (\ref{eqn:padsensitivity}). Fig.~\ref{fig:ELMperformancesensitivity} shows the normalized\footnote{The normalization is with respect to the highest value among the computed $\text{SSD}_k$.} $\text{SSD}_k$ for the five highest sensitivity meta-level features of these five machine learning methods. Note that for the HSIC-Lasso we do not use the $\text{SSD}_k$ but instead the values of the coefficient of the centered Gram matrix computed from k$^\text{th}$ feature which provides an estimate of the feature sensitivity. Recall, from Sec.~\ref{subsec:sensitivity:analysis:background}, that larger the $\text{SSD}_k$ value or higher the coefficient of the centered Gram matrix the more sensitive the \viewcount is to variations in the meta-level feature. From Fig.~\ref{fig:ELMperformancesensitivity}, the meta-level features with the highest sensitivities are: first day \viewcount, number of subscribers, contrast of the video thumbnail, Google hits, number of keywords, video category, title length, and number of upper-case letters in the title respectively. Notice that all these methods have the first day \viewcount and  number of subscribers as the most sensitive meta-level features as expected. The FFNN and Bagged MARS however do not have the contrast of the video thumbnail as the third most sensitive meta-level feature compared with the other algorithms. This results as the learning method and learning rate of each of these algorithms is different which results in differences in the meta-level feature sensitivity. However as we can see from Fig.~\ref{fig:ELMperformancesensitivity}, the \viewcount of a video is dependent on these eight meta-level features with the first day \viewcount and  number of subscribers being the most sensitive features. 

As expected, Fig.~\ref{fig:ELMperformancesensitivity} shows that if the first day \viewcount is high then the associated \viewcount 14 days after the video is posted will be high. Additionally, if there is a large number of subscribers to the channel that posted the video, then the associated \viewcount after 14 days is also expected to be large. As expected, the properties of the title and keywords also contribute to the \viewcount of the video however with less sensitivity then the thumbnail of the video. Therefore, to increase the \viewcount of a video it is vital to increase the number of subscribers, and focus on the quality of the Thumbnail used. A surprising result is that the sensitivity of the \viewcount resulting from changes in these meta-level features are negligible across the three popularity classes of videos (i.e. highly popular, popular, and unpopular as defined in Table~\ref{tab:popularity:dataset:summary}). Therefore, regardless of the expected popularity of a video, a channel owner should focus on maximizing the number of subscribers and the quality of the thumbnail to increase the associated \viewcount of a video.

\subsubsection{Predicting the \viewcount of YouTube videos}
In this section we illustrate how machine learning methods can be used to the \viewcount of a YouTube video. The machine learning methods used for prediction include: the Extreme Learning Machine~\eqref{eqn:elm}, Feed-Forward Neural Network~\cite{Bis06}, Stacked Auto-Encoder Deep Neural-Network~\cite{HZZW15,Sch15}, Elasticnet~\cite{ZH05}, Lasso, Relaxed Lasso~\cite{Mei07}, Quantile Regression with Lasso~\cite{PW15}, Conditional Inference Random Forest~\cite{HHZ06}, Boosted Generalized Additive Model~\cite{VR13,HBKSH10}, Bagged MARS using gCV Pruning~\cite{Dru97}, Generalized Linear Model with Stepwise Feature Selection using Akaike information criterion, and Spike and Slab Regression~\cite{IR05}. For each method their predictive performance and the top-five highest sensitivity meta-level features are provided. 

To perform the analysis we train each model using an identical 10-fold cross validation method with the dataset $\mathcal{D}=\{(x_i,v_i)\}_{i=1}^N$ with all the meta-level features included. The predictive performance of the machine learning methods are evaluated using the root-mean-square error (RMSE) and the R$^2$ (e.g. coefficient of determination). Note that for both training and evaluation the view count is pre-processed to be on the log scale (i.e. if the \viewcount is $10^6$, the associated label is $v_i=6$). 

The predictive performance and the top-five highest sensitivity meta-level features of the machine learning methods are provided in Table~\ref{tbl:MLperformance}. In Table~\ref{tbl:MLperformance} the meta-level feature numbers are identical to those defined in Fig.~\ref{fig:ELMperformancesensitivity}. As seen from  Table~\ref{tbl:MLperformance}, the ELM has the lowest RMSE of 0.44 which is comparable to the RMSE of the Conditional Inference Random Forest and Feed-Forward Neural Network which have 0.47 and 0.48 respectively. The R$^2$ of the ELM, Feed-Forward Neural Network and Conditional Inference Random Forest are also comparable with values of 0.77, 0.79, and 0.80. Therefore any of these methods could be used to estimate the \viewcount of a YouTube video. A key question is which of the meta-level features $x(k)$ are most sensitive between these machine learning methods. As seen from the results in Table~\ref{tbl:MLperformance} the top two most important features are the first day \viewcount and the number of subscribers, and the majority of methods suggest that the number of Google hits is also an important meta-level feature. Interestingly the Conditional Inference Random Forest, Boosted Generalized Additive Model, and the Bagged MARS using gCV Pruning do not consider the number of Google hits in the top five most sensitive features and instead use the video category. This is consistent with the result that videos in the ``Music'' category are the most viewed on YouTube, followed by ``Entertainment'' and ``People and Blogs''. Only the Bagged MARS using gCV Pruning considers the meta-level features of title length and number of upper-case letters in the title to be in the five most sensitive features compared with the other machine learning methods. This result suggests that the number of Google hits associated with the title significantly contributes to the video's popularity however the \viewcount is not very sensitive to the specific length and number of upper-case letters in the title. Therefore, when performing meta-level feature optimization for a video a user should focus on the meta-level features of: first day \viewcount, number of subscribers, contrast of the video thumbnail, Google hits, number of keywords, and video category.

\begin{table}[h!]
	\centering
	\caption{Performance and Feature Sensitivity}
	\resizebox{\columnwidth}{!}{%
	\begin{tabular}{|l|c|c|c|c|c|c|c|}
		\toprule
		{\bf Method} &  {\bf RMSE} & {\bf R$^2$} & \multicolumn{5}{|c|}{\bf Features $x(k)$} \\\hline
		Extreme Learning Machine & {\bf 0.44} & 0.77 & 1 & 2 & 3 & 4 & 5 \\
		Feed-Forward Neural Network & 0.48 & 0.79 & 2 & 1 & 5 & 3 & 6 \\
		Stacked Auto-Encoder DNN & 0.59 & 0.66 & 1 & 2 & 3 & 4 & 5 \\
		Elasticnet & 0.57 & 0.64 & 1 & 2 & 3 & 4 & 5 \\
		Lasso & 0.53 & 0.66 & 1 & 2 & 3 & 4 & 5 \\
		Relaxed Lasso & 1.14 & 0.64 & 1 & 2 & 3 & 4 & 5 \\
		Quantile Regression with Lasso  & 0.60 & 0.62 & 1 & 2 & 3 & 4 & 5 \\
		CI Random Forest  & 0.47 & {\bf 0.80} & 1 & 2 & 6 & 4 & 5 \\
		Boosted GAM  & 0.50 & 0.77 & 1 & 2 & 6 & 4 & 5 \\
		Bagged MARS & 0.50 & 0.77 & 1 & 2 & 6 & 7 & 8 \\
		GLM with Feature Selection & 0.53 & 0.67 & 1 & 2 & 3 & 4 & 5 \\
		Spike and Slab Regression & 0.53 & 0.67 & 1 & 2 & 3 & 4 & 5 \\
		\bottomrule
	\end{tabular}%
	}
	\label{tbl:MLperformance}
\end{table}

\begin{figure*}[h!]
  \centering
\subfigure[]{\label{fig:ELMperformancesensitivityBA} \includegraphics[angle=0,clip=1,trim=40 0 40 0,width=3.2in]{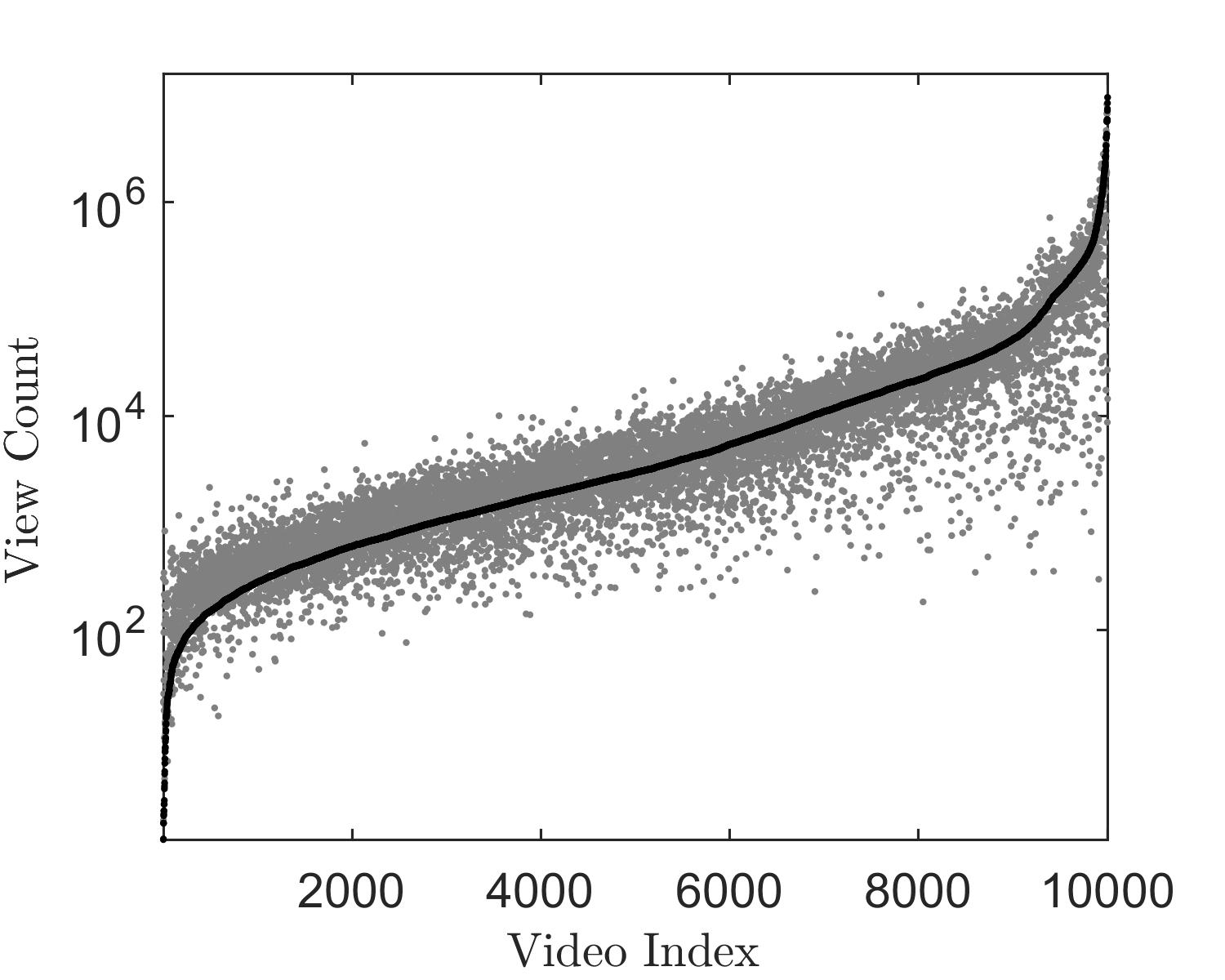}} 
\subfigure[]{\label{fig:ELMperformancesensitivityBB} \includegraphics[angle=0,clip=1,trim=40 0 40 0,width=3.2in]{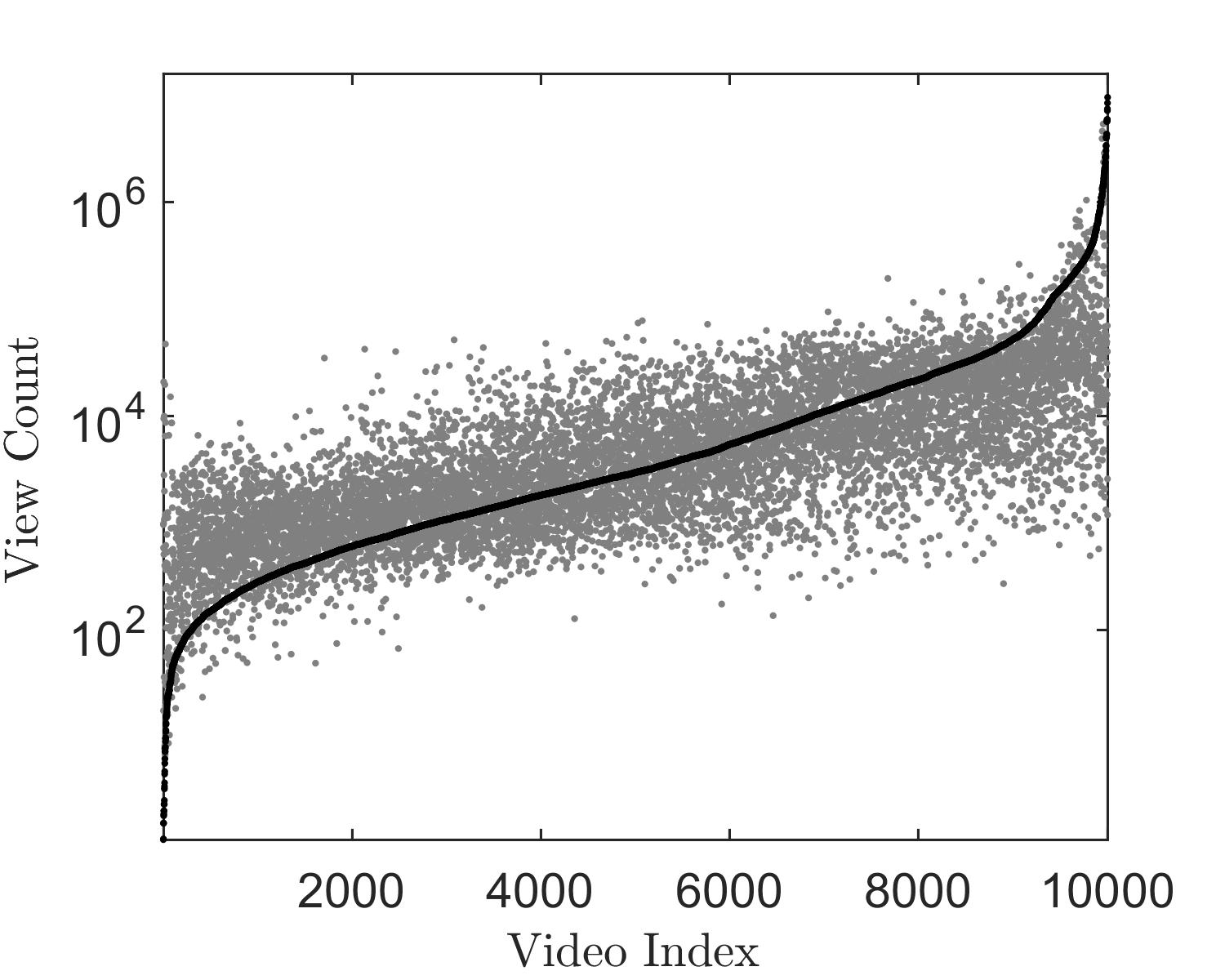}} 
\caption[ELM predictive accuracy]{Predictive \viewcount using an ELM with the actual \viewcount (black dots) and predicted \viewcount indicated by the (gray dots). Fig.~\ref{fig:ELMperformancesensitivityBA} illustrates the results for a trained ELM (\ref{eqn:elm}) using all 29 meta-level features defined in Sec.~\ref{subsec:ELMsensitivity}. Fig.~\ref{fig:ELMperformancesensitivityBB} illustrates the results for a trained ELM (\ref{eqn:elm}) using the 28 meta-level features (first day view count removed from the 29 meta-level features defined in Sec.~\ref{subsec:ELMsensitivity}).}
\label{fig:ELMperformancesensitivityB} 
\end{figure*}

To estimate the \viewcount of an unpublished video (a video that is about to be posted for the first time) we can not utilize the most sensitive meta-level feature of the machine learning algorithms which is the first day \viewcount. Is it still possible to estimate the \viewcount with the remaining meta-level features? To answer this question we compare the performance of the ELM using the 28 meta-level features with the \viewcount on the first day removed. Fig.~\ref{fig:ELMperformancesensitivityBA} shows the predicted \viewcount of the ELM trained using 29 meta-level features, and Fig.~\ref{fig:ELMperformancesensitivityBB} shows the predicted \viewcount using the 28 meta-level features. 
As expected, Fig.~\ref{fig:ELMperformancesensitivityB} illustrates that the predictive accuracy of the ELM decreases if the \viewcount on the first day is removed. Though there is a drop in the predictive accuracy of the ELM trained using the 28 meta-level features, it still contains sufficient predictive accuracy to aid in the selection of the meta-level features to increase the \viewcount of a video. Note that similar performance results are obtained for the Feed-Forward Neural Network and Conditional Inference Random Forest when performing the prediction with the first day \viewcount removed. Therefore these prediction methods can be used to optimize the meta-level features of unpublished videos where the optimization can focus on the meta-level features of: number of subscribers, contrast of the video thumbnail, Google hits, number of keywords, and video category.


\subsection{Sensitivity to meta-level optimization}
\label{subsec:sensitivity:metalevel:optimization}
Sec.~\ref{subsec:ELMsensitivity}, described how meta-level features (e.g. number of subscribers) can be used to estimate the popularity of a video. 
In this section, we analyze how changing \metalevel features, after a video is posted, impacts the user engagement of the video. 
Meta-level data plays a significant role in the discovery of content, through YouTube search, and in video recommendation, through the YouTube related videos. 
Hence, ``optimizing'' the \metalevel data to enhance the discoverability and user engagement of videos is of significant importance to content providers. 
Therefore, in this section, we study how optimizing the title, thumbnail or keywords affect the view count of YouTube videos. 

To perform the analysis we utilize the dataset (see Table~\ref{tab:optimization:dataset:summary} in the Appendix), and remove any time-sensitive videos. Time-sensitive videos are those videos that are relevant for a short period of time and the popularity of such videos cannot be improved by optimization. We removed the following two time-sensitive categories of videos:  ``politics'' and ``movies and trailers''. In addition, we removed videos (from other categories) which contained the following keywords in their video meta-data: ``holiday'', ``movie'', or ``trailers''. For example, holiday videos are not watched frequently during off-holiday times. 

Let $\hat{\tau}_i$ be the time at which the \metalevel optimization was performed on video $i$ and let $s_i$, denote the corresponding sensitivity. 
We characterize the sensitivity to \metalevel optimization as follows: 
\begin{equation}
	s_i = \frac{
		\left(
		\sum_{t =\hat{\tau}_i}^{\hat{\tau}_i+6}
			v_i(t)
			\right)/7
		}
		{
		\left(
		\sum_{t =
			\hat{\tau}_i-6}^{\hat{\tau}_i}
			v_i(t) 
			\right)/7}
	\label{eqn:sensitivity:optimization:characterize}
\end{equation}
The numerator of~\eqref{eqn:sensitivity:optimization:characterize} is the mean value of the view count $7$ days after optimization. Similarly, the denominator of~\eqref{eqn:sensitivity:optimization:characterize} is the mean value of the view count $7$ days before optimization. 
The results are provided in Table~\ref{tab:sensitivity:optimization} for optimization of the title, thumbnail, and keywords. 
\begin{table}[!h]
	\centering
	\begin{tabular}{c|c}
		\toprule
		Optimization & Fraction of Videos with increased popularity \\
		\midrule
		Title change &  0.52 \\
		Thumbnail change &  0.533 \\
		Keyword change &  0.50 \\
		No change\footnotemark &  0.35 \\
		\bottomrule
	\end{tabular}
	\caption{Sensitivity to Meta-Level Optimization. The table shows than in more than $50\%$ the videos, \metalevel optimization resulted in an increase in the popularity of the video. \label{tab:sensitivity:optimization}}
\end{table} \footnotetext{``No change'' was obtained by randomly selecting $10^4$ videos which performed no optimization and evaluating $s_i$ $3$ months from the date of posting the video. }
As shown in Table~\ref{tab:sensitivity:optimization}, at least half of the optimizations resulted in an increase in the popularity of the video. In addition, compared to videos with no optimization, the \metalevel optimization improves the probability of increased popularity by $45\%$. 
This is consistent with YouTube and BBTV recommendation to optimize \metalevel features to increase user engagement.  
However, some class of videos benefit from optimizing meta-data much more than others. 
The effect may be due to small user channels, which have limited number of videos and subscribers, gain by optimizing the meta-level data of the video compared to hugely popular channels such as Sony or CNN.  
The highly popular channel (e.g.\ Sony or CNN) upload videos frequently (even multiple times daily), so video content becomes irrelevant quickly. 
The question of which class of users gain by optimizing the meta level features of the video is part of our ongoing research. 

Table~\ref{tab:sensitivity:traffic} summarizes the impact of various meta-level changes on the three major sources of YouTube traffic, i.e. YouTube search\footnote{Video views from YouTube search results}, YouTube promoted\footnote{Video views from an unpaid YouTube promotion} and traffic from related videos\footnote{Video views from a related video listing on another video watch page}. For those videos where \metalevel optimization increased the popularity (the ratio of the mean value of the views after and before optimization is higher than one), we computed the sensitivity for various traffic sources as in~\eqref{eqn:sensitivity:optimization:characterize}. Table~\ref{tab:sensitivity:traffic} summarizes the median statistics of the ratio of the traffic sources before and after optimization. 
\begin{table}[!h]
\begin{minipage}{\columnwidth}
		\begin{tabular*}{\textwidth}{@{\hspace{\tabcolsep}\extracolsep{\fill}}cccc}
		\toprule
		Optimization & Related &  Promoted & Search\\
		\midrule
		Title change &   $1.13$ & NA\footnote{Not enough data available: A binomial test to check for the true hypothesis with $95\%$ confidence interval requires that the sample size, $n$, should be at least $\left(\frac{1.96}{0.04}\right)^2 p (1-p)$. With $p = 0.5$, $n > 600$. \label{foot:data:na}} & $\mathbf{1.24}$ \\
		Thumbnail change &  $\mathbf{1.20}$ & NA\footref{foot:data:na} & $1.125$\\
		Keyword change & $\mathbf{1.10}$ & $\mathbf{1.16}$ & $1$  \\
		\bottomrule
		\end{tabular*}
\end{minipage}%
	\caption{Sensitivity of various traffic sources to \metalevel optimization, for videos with increased popularity. The title optimization resulted in significant improvement (approximately $25\%$) from the YouTube search. Similarly, thumbnail optimization improved traffic from the related videos and keyword optimization resulted in increased traffic from related and promoted videos. }
	\label{tab:sensitivity:traffic}
\end{table}
The title optimization resulted in significant improvement (approximately $25\%$) from the YouTube search. Similarly, thumbnail optimization improved traffic from the related videos and keyword optimization resulted in increased traffic from related and promoted videos. 

{\em Summary}: This section studied the sensitivity of \viewcount with respect to \metalevel optimization. 
The main finding is that \metalevel optimization increased the popularity of video in the majority of cases. 
In addition, we found that optimizing the title improved traffic from YouTube search. Similarly, thumbnail optimization improved traffic from the related videos and keyword optimization resulted in increased traffic from related and promoted videos. 
\section{Social interaction of the channel with YouTube users}
\label{sec:social:dynamics:YouTube}
In this section, we use time series analysis methods to determine how the social interaction of a YouTube channel with its viewers affects the \viewcount dynamics. 
This section is organized as follows. 
Sec.~\ref{sec:causal:relation:YouTube}, characterizes the causal relationship between the subscribers and \viewcounts of a channel using Granger causality test. 
In Sec.~\ref{sec:scheduling:YouTube}, we investigate how the popularity of the channel is affected by the scheduling dynamics of the channel. 
When channels deviate from a regular upload schedule, the \viewcounts and the comment count of the channel increase. 
In Sec.~\ref{sec:seperate:virality:migration}, we address the problem of separating the \viewcount dynamics due to virality (viewcount resulting from subscribers) and migration (views from non-subscribers) and exogenous events using a generalized Gompertz model. 
Finally, Sec.~\ref{sec:playlist:dynamics}, we studies the effect of video playlists on the view count. 
The main conclusion outlined in Sec.~\ref{sec:playlist:dynamics} is that the dynamics of the \viewcount in a playlist is highly correlated and the effects of ``migration'' causes the \viewcount of videos to decrease even with an increase in the subscriber count. 

\subsection{Causality between subscribers and \viewcounts in YouTube}
\label{sec:causal:relation:YouTube}
\begin{figure}[h]
	\centering
	\input{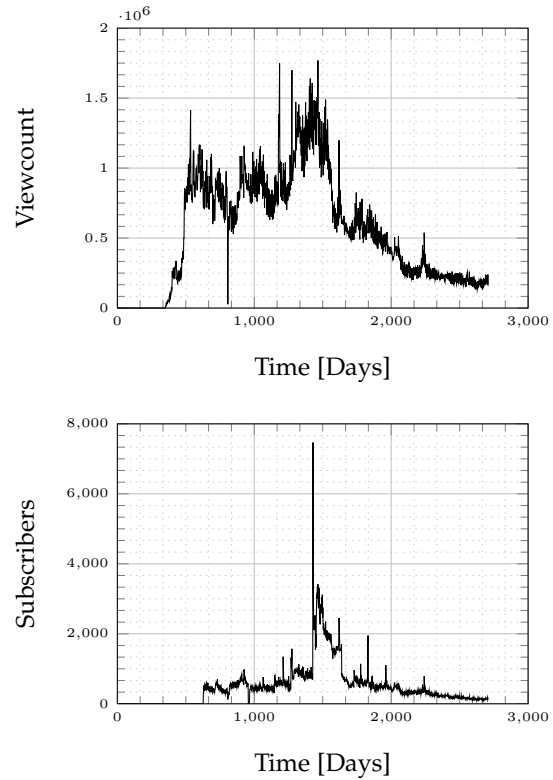}
	\caption{Viewcount and subscribers for the popular movie trailer channel: VISOTrailers. The Granger causality test for view counts ``Granger causes'' subscriber count is true with a p-value of $5\times 10^{-8}$. }
	\label{fig:Channel6:sub:view}
\end{figure}

In this section the goal is to detect the causal relationship between subscriber and viewer counts and how it can be used to estimate the next day subscriber count of a channel. The results are of interest for measuring the popularity of a YouTube channel. 
Fig.~\ref{fig:Channel6:sub:view} displays the subscriber and view count dynamics of a popular movie trailer channel in YouTube. 
It is clear from Fig.~\ref{fig:Channel6:sub:view} that the subscribers ``spike'' with a corresponding ``spike'' in the view count. 
In this section we model this causal relationship of the subscribers and \viewcount using the Granger causality test from the econometric literature~\cite{granger1969investigating}. 

The main idea of Granger causality is that if the value(s) of a lagged time-series can be used to predict another time-series, then the lagged time-series is said to ``Granger cause'' the predicted time-series. 
%
To formalize the Granger causality model, let $s^j(t)$ denote the number of subscribers to a channel $j$ on day $t$, and $v_i^j(t)$ the corresponding view count for a video $i$ on channel $j$ on day $t$. The total number of videos in a channel on day $t$ is denoted by $\mathcal{I}(t)$. Define, 
\begin{equation}
	\hat{v}^j(t) = \sum_{i=1}^{\mathcal{I}(t)}v_i^j(t),
\end{equation}
as the total view count of channel $j$ at time $t$. 
The Granger causality test involves testing if the coefficients $b_i$ are non-zero in the following equation which models the relationship between subscribers and view counts: 
\begin{equation}
	s^j(t) = \sum_{k = 1}^{n_s}a_k^j s^j(t-k) + \sum_{i=k}^{n_v}b_k^j \hat{v}^j(t-k) + \varepsilon^j(t),
	\label{eqn:arma:model}
\end{equation}
where $\varepsilon^j(t)$ represents normal white noise for channel $j$ at time $t$. The parameters $\{a_i^j\}_{\{i=1,\dots,n_s\}}$ and $\{b_i^j\}_{\{i=1,\dots,n_v\}}$ are the coefficients of the AR model in~\eqref{eqn:arma:model} for channel $j$, with $n_s$ and $n_v$ denoting the lags for the subscriber and view counts time series respectively. 
If the time-series $\mathcal{D}^j=\{s^j(t),\hat{v}^j(t)\}_{t\in\{1,\dots,T\}}$ of a channel $j$ fits the model~\eqref{eqn:arma:model}, then we can test for a causal relationship between subscribers and view count. 
In equation~\eqref{eqn:arma:model}, it is assumed that $|a_i| < 1$, $|b_i| < 1$ for stationarity. The causal relationship can be formulated as a hypothesis testing problem as follows:  
\begin{equation}
	H_0: b_1 = \dots = b_{n_v}=0 \text{ vs. } H_1: \text{Atleast one } b_i \neq 0. 
	\label{eqn:hypothesis}
\end{equation}
The rejection of the null hypothesis, $H_0$, implies that there is a causal relationship between subscriber and view counts.

First, we use Box-Ljung test~\cite{LB78} is to evaluate the quality of the model~\eqref{eqn:arma:model} for the given dataset $\mathcal{D}^j$. If satisfied, then the Granger causality hypothesis~\eqref{eqn:hypothesis} is evaluated using the Wald test~\cite{Wald73}. If both hypothesis tests pass then we can conclude that the time series $\mathcal{D}^j$ satisfies Granger causality--that is, the previous day subscriber and view count have a causal relationship with the current subscriber count. 

A key question prior to performing the Granger causality test is what percentage of videos in the YouTube dataset (Appendix) satisfy the AR model in~\eqref{eqn:arma:model}. To perform this analysis we apply the Box-Ljung test with a confidence of 0.95 (p-value = $0.05$). 
First, we need to select $n_s$ and $n_v$, the number of lags for the subscribers and view count time series. For $n_s=n_v = 1$, we found that only $20\%$ of the channels satisfy the model~\eqref{eqn:arma:model}. When $n_s$ and $n_v$ are increased to $2$, the number of channels satisfying the model increases to $63\%$.  For $n_s=n_v = 3$, we found that $91\%$ of the channels satisfy the model~\eqref{eqn:arma:model}, with a confidence of $0.95$ (p-value = $0.05$). Hence, in the below analysis we select $n_s=n_v = 3$. It is interesting to note that the mean value of coefficients $b_i$ decrease as $i$ increases indicating that older view counts have less influence on the subscriber count. Similar results also hold for the coefficients $a_i$. Hence, as expected, the previous day subscriber count and the previous day view count most influence the current subscriber count.  

The next key question is does their exist a causal relationship between the subscriber dynamics and the view count dynamics.
This is modeled using the hypothesis in~\eqref{eqn:hypothesis}. 
To test~\eqref{eqn:hypothesis} we use the Wald test with a confidence of $0.95$ (p-value = $0.05$) and found that approximately $55\%$ of the channels satisfy the hypothesis. For approximately $55\%$ of the channels that satisfy the AR model~\eqref{eqn:arma:model}, the view count ``Granger causes'' the current subscriber count. Interestingly, if different channel categories are accounted for then the percentage of channels that satisfy Granger causality vary widely as illustrated in Table~\ref{tab:category:causality}. For example, $80\%$ of the Entertainment channels satisfy Granger causality while only $40\%$ of the Food channels satisfy Granger causality. These results illustrate the importance of channel owners to not only maximize their subscriber count, but to also upload new videos or increase the views of old videos to increase their channels popularity (i.e. via increasing their subscriber count). Additionally, from our analysis in Sec.\ref{sec:sensitivity:analysis:elm} which illustrates that the view count of a posted video is sensitive to the number of subscribers of the channel, increasing the number of subscribers will also increase the view count of videos that are uploaded by the channel owners. 
\begin{table}[!h]
\begin{minipage}{\columnwidth} 
	\begin{center}
	\begin{tabular}{c|c}
		\toprule
		Category\footnote{YouTube assigns a category to videos, rather than channels. The category of the channel was obtained as the majority of the category of all the videos uploaded by the channel. } & Fraction \\
		\midrule
		Gaming & 0.60 \\
		Entertainment & 0.80 \\
		Food & 0.40 \\
		Sports & 0.67\\
		\bottomrule
	\end{tabular}
\end{center}
\end{minipage}%
	\caption{Fraction of channels satisfying the hypothesis: View count ``Granger causes'' subscriber count, split according to category.}
	\label{tab:category:causality}
\end{table}
\subsection{Scheduling dynamics in YouTube}
\label{sec:scheduling:YouTube}
In this section, we investigate the scheduling dynamics of YouTube channels. 
We find the interesting property that for popular gaming YouTube channels with a dominant upload schedule, deviating from the schedule increases the views and the comment counts of the channel. 

Creator Academy\footnote{YouTube website for helping with channels} in their best practice section recommends to upload videos on a regular schedule to get repeat views. 
The reason for a regular upload schedule is to increase the user engagement and to rank higher in the YouTube recommendation list. 
However, we show in this section that going ``off the schedule'' can be beneficial for a gaming YouTube channel, with a regular upload schedule, in terms of the number of views and the number of comments.  

From the dataset, we `filtered out' video channels with a \emph{dominant} upload schedule, as follows: The dominant upload schedule was identified by taking the periodogram of the upload times of the channel and then comparing the highest value to the next highest value. 
If the ratio defined above is greater than $2$, we say that the channel has a dominant upload schedule. 
From the dataset containing $25$ thousand channels, only $6500$ channels contain a dominant upload schedule. 
Some channels, particularly those that contain high amounts of copied videos such as trailers, movie/TV snippets upload videos on a daily basis. 
These have been removed from the above analysis. The expectation is that by doing so we concentrate on those channels that contain only user generated content. 

We found that channels with gaming content account for $75\%$ of the $6500$ channels with a dominant upload schedule\footnote{This could also be due to the fact gaming videos account for 70\% of the videos in the dataset.} and the main tags associated with the videos were: ``game'', ``gameplay'' and ``videogame''\footnote{We used a topic model to obtain the main tags. }.  
We computed the average views when the channel goes off the schedule and found that on an average when the channel goes off schedule the channel gains views $97\%$ of the time and the channel gains comments $68\%$ of the time.  
This suggests that channels with ``gameplay'' content have periodic upload schedule and benefit from going off the schedule.  
\subsection{Modeling the View count Dynamics of Videos with Exogenous Events}
\label{sec:seperate:virality:migration}
\label{sec:detection:exogenous}
Several time-series analysis methods have been employed in the literature to model the \viewcount dynamics of YouTube videos. These include ARMA time series models~\cite{GCM11}, multivariate linear regression models~\cite{PAG13}, hidden Markov models~\cite{JMYLH14}, normal distribution fitting~\cite{FBA11}, and parametric model fitting~\cite{RAEJLP14,REJAL15}. Though all these models provide an estimate of the \viewcount dynamics of videos, we are interested in segmenting \viewcount dynamics of a video resulting from subscribers, non-subscribers and exogenous events.  Exogenous events are due to video promotion on other social networking platform such as Facebook or the video being referenced by a popular news organization or celebrity on Twitter. This is motivated by two reasons. First, removing \viewcount dynamics due to exogenous events provides an accurate estimate of sensitivity of \metalevel features in Sec.~\ref{sec:sensitivity:analysis:elm}. Second, extracting the \viewcount resulting from exogenous events gives an estimate of the efficiency of video promotion.  

The view count dynamics of popular videos in YouTube typically show an initial viral behaviour, due to subscribers watching the content, and then a linear growth resulting from non-subscribers. The linear growth is due to new users migrating from other channels or due to interested users discovering the content either through search or recommendations (we call this phenomenon \emph{migration} similar to~\cite{RAEJLP14}). Hence, without exogenous events, the \viewcount dynamics of a video due to subscribers and non-subscribers can be estimated using piecewise linear and non-linear segments.  In~\cite{RAEJLP14}, it is shown that a Gompertz time series model can be modeled the \viewcount dynamics from subscribers and non-subscribers, if no exogenous events are present. In this paper, we generalize the model in~\cite{RAEJLP14} to account for views from exogenous events. It should be noted that classical change-point detection methods~\cite{TNB14} cannot be used here as the underlying distribution generating the \viewcount is unknown. 

To account for the \viewcount dynamics introduced from exogenous events we use the generalized Gompertz model given by: 
\begin{align}
\begin{aligned}
\bar{v}_i(t) &= \sum_{k=0}^{K_\text{max}} w^k_i(t)u(t-t_k), \\
w_i^k(t) &= M_k\left(1-e^{-\eta_k\left(e^{b_k\left(t-t_k\right)}-1\right)}\right)+c_k(t-t_k),
\end{aligned}
\label{eqn:gompertz}
\end{align}
where $\bar{v}_i(t)$ is the total \viewcount for video $i$ at time $t$, $u(\cdot)$ is the unit step function, $t_0$ is the time the video was uploaded, $t_k$ with $k\in\{1,\dots,K_\text{max}\}$ are the times associated with the $K_\text{max}$ exogenous events, and $w^k_i(t)$ are Gompertz models which account for the \viewcount dynamics from uploading the video and from the exogenous events. In total there are $K_\text{max}+1$ Gompertz models with each having parameters $t_k, M_k, \eta_k, b_k$. $M_k$ is the maximum number of requests not including migration for an exogenous event at $t_k$, $\eta_k$ and $b_k$ model the initial growth dynamics from event $t_k$, and $c_k$ accounts for the migration of other users to the video. In (\ref{eqn:gompertz}) the parameters $\{M_k,\eta_k,b_k\}_{k=0}$ are associated with the subscriber views when the video is initially posted, the parameters $\{t_k,M_k,\eta_k,b_k\}_{k=1}^{K_\text{max}}$ are associated with views introduced from exogenous events, and the views introduced from migration are given by $\{c_k\}_{k=0}^{K_\text{max}}$. Each Gompertz model (\ref{eqn:gompertz}) captures the initial viral growth when the video is initially available to users, followed by a linearly increasing growth resulting from user migration to the video. 

\begin{figure}[h]
	\centering
	\includegraphics[angle=0,width=1.5\figurewidth]{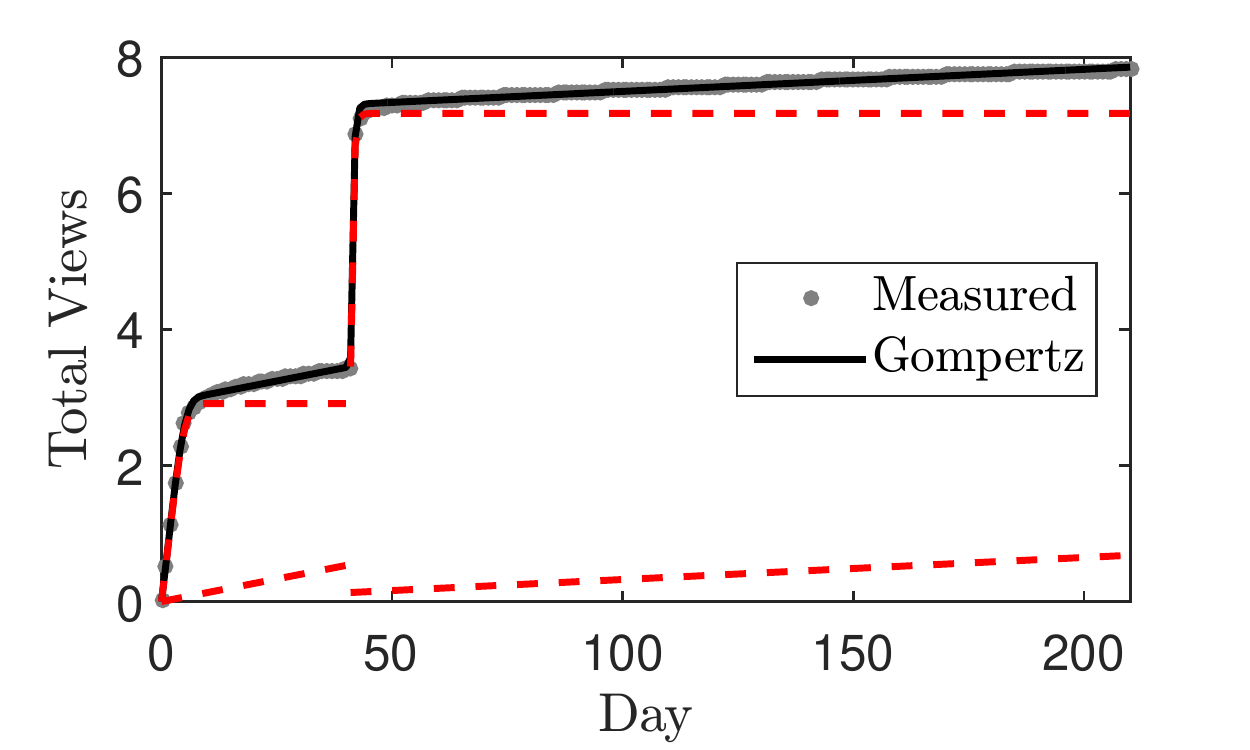}
	\caption{Due to an exogenous event on day $41$, there is a sudden increase in the number of views. The total view count fitted by the Gompertz model $\bar{v}_i(t)$ in~\eqref{eqn:gompertz} is shown in black with the virality (exponential) and migration (linear) illustrated by the dotted red. }
	\label{fig:ExoFitExoGompertz}
\end{figure}

The parameters $\theta_i=\{a_k,t_k,M_k,\eta_k,b_k,c_k\}_{k=0}^{K_\text{max}}$ in (\ref{eqn:gompertz}) can be estimated by solving the following mixed-integer non-linear program:
\begin{align}
\theta_i&\in\operatorname{arg\ min}\Big\{\sum_{t=0}^{T_i}\big(\bar{v}_i(t)-v_i(t)\big)^2+\lambda K\Big\} \nonumber\\
&K=\sum_{k=0}^{K_\text{max}}a_k, \quad\quad a_k\in\{0,1\} \quad\quad k\in\{0,\dots,K_\text{max}\},
\label{eqn:gompertzMINLP}
\end{align}
with $T_i$ the time index of the last recorded views of video $v_i$, and $a_k$ a binary variable equal to $1$ if an exogenous event is present at $t_k$. Note that (\ref{eqn:gompertzMINLP}) is a difficult optimization problem as the objective is non-convex as a result of the binary variables $a_k$~\cite{BL12}. In the YouTube social network when an exogenous event occurs this causes a large and sudden increase in the number of views, however as seen in Fig.~\ref{fig:ExoFitExoGompertz}, a few days after the exogenous event occurs the views only result from migration (i.e. linear increase in total views). Assuming that each exogenous event is followed by a linear increase in views we can estimate the total number of exogenous events $K_\text{max}$ present in a given time-series by first using a segmented linear regression method, and then counting the number of segments of connected linear segments with a slope less then $c_\text{max}$. The parameter $c_\text{max}$ is the maximum slope for the views to be considered to result from viewer migration. Plugging $K_\text{max}$ into (\ref{eqn:gompertzMINLP}) results in the optimization of a non-linear program for the unknowns $\{t_k,M_k,\eta_k,b_k,c_k\}_{k=0}^{K_\text{max}}$. This optimization problem can be solved using sequential quadratic programming techniques~\cite{Ber99}. 

To illustrate how the Gompertz model~\eqref{eqn:gompertz} can be used to detect for exogenous events, we apply (\ref{eqn:gompertz}) to the \viewcount dynamics of a video that only contains a single exogenous event. Fig.~\ref{fig:ExoFitExoGompertz} displays the total \viewcounts of a video where an exogenous event occurs at time $t=41$ (i.e.\ $t_1 = 41$ in~\eqref{eqn:gompertz}) days after the video is posted\footnote{Due to privacy reasons, we cannot detail the specific event. Some of the reasons for the sudden increase in the popularity of the video include: Another user on YouTube mentioning the video, this will encourage viewers from that channel to view the video, resulting in a sudden increase in the number of views. Another possibility is that the channel owner or a YouTube Partner like BBTV did significant promotional initiatives on other social media sites such as Twitter, Facebook, etc. to promote the channel or video.}. 
The initial increase in views for the video for $t\leq7$ days results from the $2910$ subscribers of the channel viewing the video. For $7\leq t \leq 41$, other users that are not subscribed to the channel migrate to view the video at an approximately constant rate of $13$ views/day. At $t=41$, an exogenous event occurs causing an increase in the views per day. The difference in viewers, resulting from the exogenous event, is $7174$. For $t\geq 43$, the views result primarily from the migration of users to approximately $2$ views/day. Hence, using the generalized Gompertz model (\ref{eqn:gompertz}) we can differentiate between subscriber views, views caused by exogenous events, and views caused by migration.
\subsection{Video Playthrough Dynamics}
\label{sec:playlist:dynamics}
\begin{figure}[h!]
  \centering
  \subfigure[Actual and predicted \viewcount of playthrough. We plot the \nth{1}, \nth{5}, \nth{10}, \nth{15}, \nth{20} and \nth{25} video from the playlist containing $25$ videos. In the legend, \emph{Exp} and \emph{Pred} corresponds to the actual and the predicted value using~\eqref{eqn:gompertz}, respectively. Figure shows that the view counts decreases for subsequent videos in the playlist. ]{\label{fig:numpred}\includegraphics[scale=0.5]{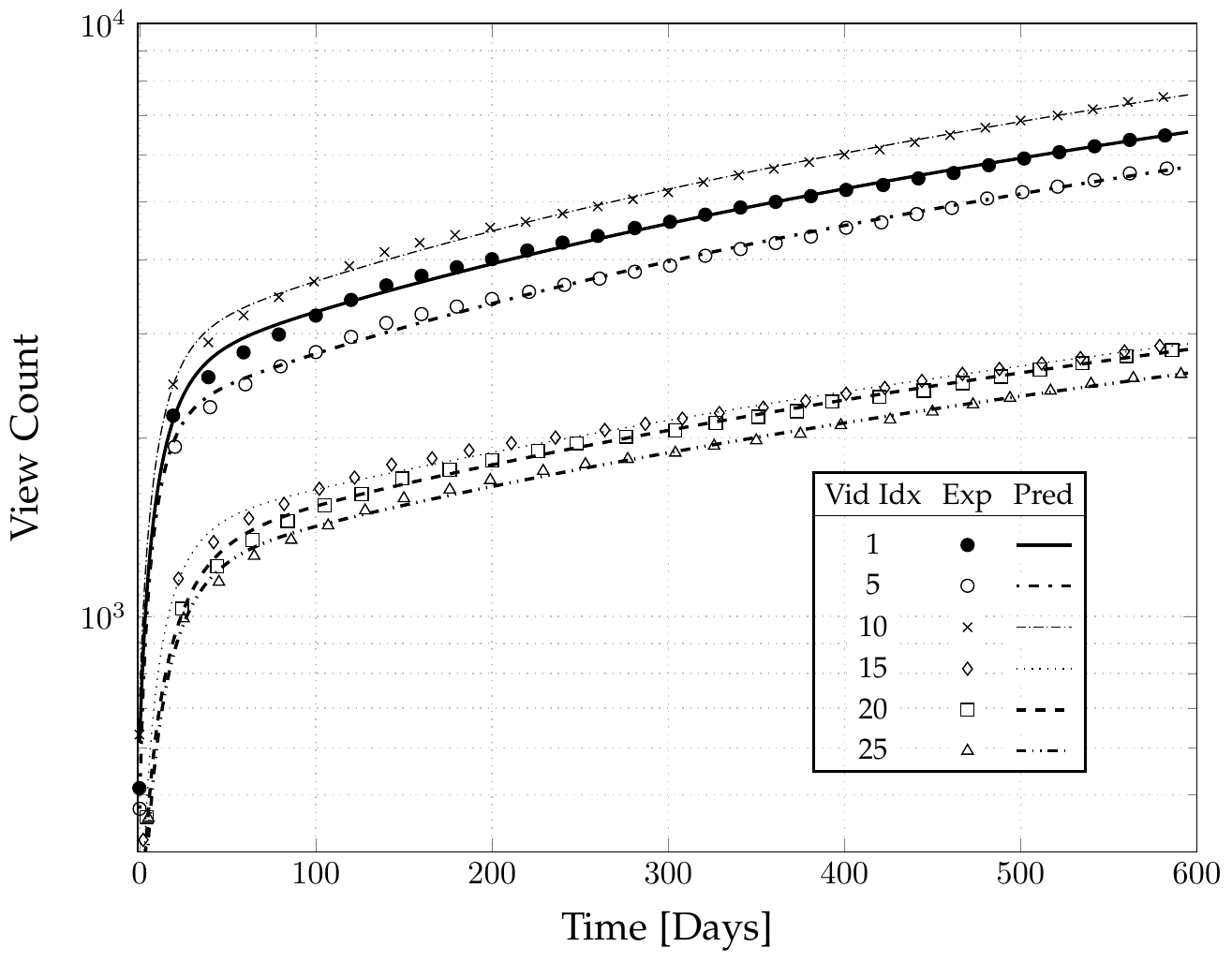}} \\
  \subfigure[The virality rate specifies the early views due to subscribers, and the migration rate (in units of views/1000 days) specifies the subsequent linear growth due to non-subscribers. ]{\label{fig:earlylaterelation}\includegraphics[scale=0.5]{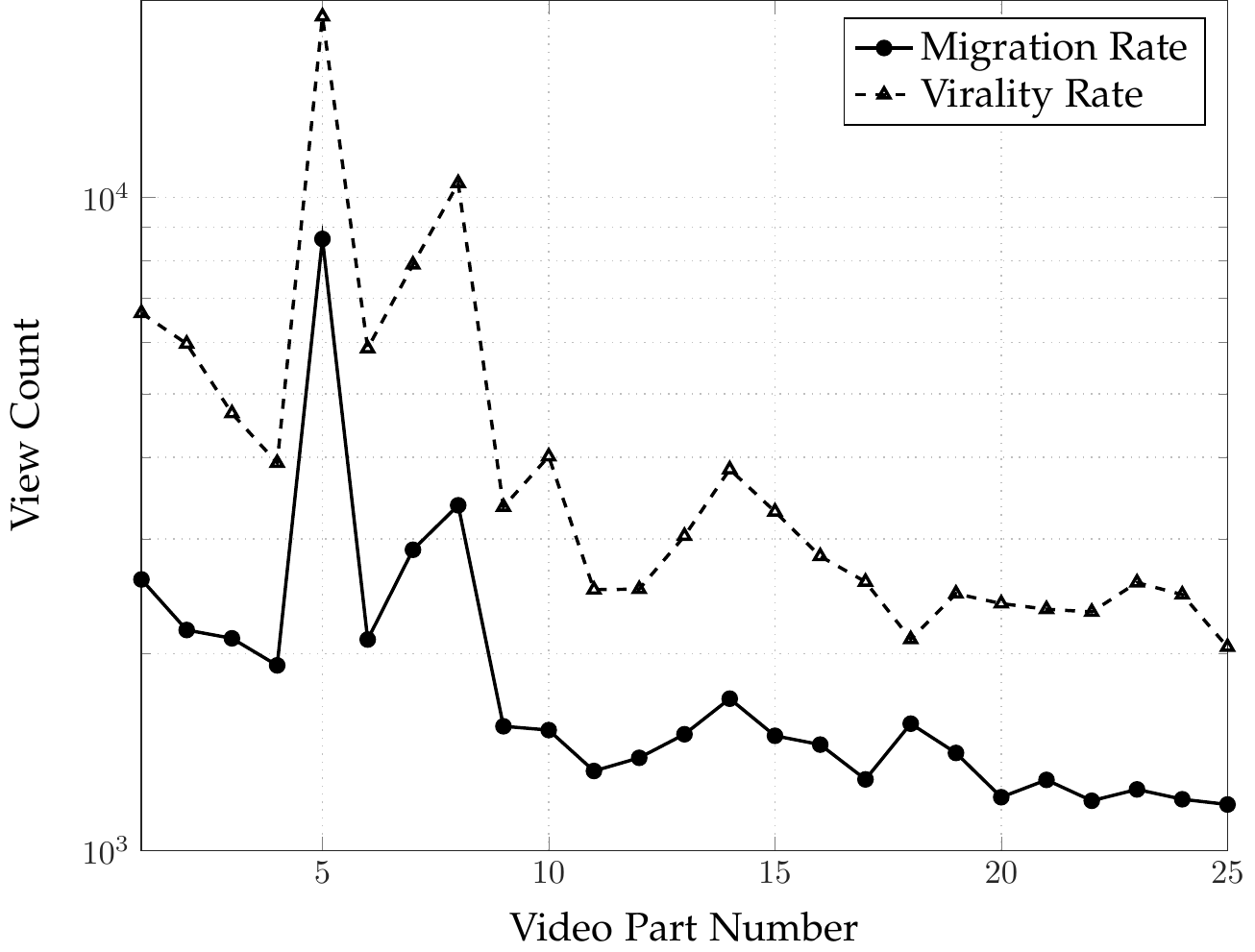}}
\vspace{-10pt}
\caption{Actual and predicted \viewcount of a playthrough containing 25 YouTube videos for the game ``BioShock Infinite''. The predictions are computed by fitting a modified Gompertz model (\ref{eqn:gompertz}) to the measured \viewcounts for each video in the playthrough.}
\label{fig:BioInfPlaythrough}
\vspace{-10pt}
\end{figure}

One of the most popular sequences of YouTube videos is the video game ``playthrough''. A video game playthrough is a set of videos for which each video has a relaxed and casual focus on the game that is being played and typically contains commentary from the user presenting the playthrough. Unlike YouTube channels such as CNN, BBC, and CBC in which each new video can be considered independent from the others, in a video playthrough the future \viewcount of videos are influenced by the previously posted videos in the playthrough. 
To illustrate this effect we consider a video playthrough for the game ``BioShock Infinite''--a popular video game released in 2013. 
The channel, popular for hosting such video playthroughs, contains close to $4500$ videos and $180$ video playthroughs. 
The channel is highly popular and has garnered a combined view count close to $100$ million views with $150$ thousand subscribers over a period of $3$ years. 
Fig.~\ref{fig:BioInfPlaythrough} illustrates that the early view count dynamics are highly correlated with the \viewcount dynamics of future videos. Both the short term \viewcount and long term migration of future videos in the playthrough decrease after the initial video in the playthrough is posted. This results for two reasons, either the viewers purchase the game, or the viewers leave as the subsequent playthroughs become repetitive as a result of game quality or video commentary quality. A unique effect with video playthroughs is that though the number of subscribers to the channel hosting the videos in Fig.~\ref{fig:BioInfPlaythrough} increases over the 600 day period, the linear migration is still maintained after the initial 50 days after the playthrough is published. Additionally, the slope of the migration is related to the early total \viewcounts as illustrated in Fig.~\ref{fig:earlylaterelation}. 
\setlength{\figurewidth}{4.52083in}
\setlength{\figureheight}{3.53854in}

\section{Conclusion}
\label{sec:conclusion}
In this paper, we conducted a data-driven study of YouTube based on a large dataset (see Appendix for details).  First, by using several machine learning methods, we investigated the sensitivity of the videos meta-level features on the view counts of videos. It was found that the most important meta-level features include: first day \viewcount, number of subscribers, contrast of the video thumbnail, Google hits, number of keywords, video category, title length, and number of upper-case letters in the title respectively. Additionally, optimizing the meta-data after the video is posted improves the popularity of the video. The social dynamics (the interaction of the channel) also affects the popularity of the channel. Using the Granger causality test, we showed that the \viewcount has a casual effect on the subscriber count of the channel. A generalized Gompertz model was also presented which can allow the classification of a videos view count dynamics which results from subscribers, migration, and exogenous events. This is an important model as it allows the views to be categorized as resulting from the video or from exogenous events which bring viewers to the video. The final result of the paper was to study the upload scheduling dynamics of gaming channels in YouTube. It was found that going ``off schedule'' can actually increase the popularity of a channel. Our conclusions are based on the BBTV dataset. Extrapolating these results to other YouTube datasets is an important problem worth addressing in future work. Another extension of the current work could involve studying the effect of video characteristics on different traffic sources, for example the affect of tweets or posts of videos on Twitter or Facebook.

%
%



\appendix
\section*{Description of YouTube Dataset}
This paper uses the dataset provided by BBTV. 
The dataset contains daily samples of metadata of YouTube videos on the BBTV platform from April, 2007 to May, 2015, and has a size of around $200$ gigabytes. 
The dataset contains around $6$ million videos spread over $25$ thousand channels. 
Table~\ref{tab:dataset:summary} shows the statistics summary of the videos present in the dataset. 
\begin{table}[h!]
	\centering
	\caption{Dataset summary}
	\begin{tabular}{c|c}
		\toprule
		Videos &  $6$ million\\
		Channels &  $26$ thousand\\
		Average number of videos (per channel)& 250 \\
		Average age of videos & 275 days\\
		Average number of views  (per video) & $10$ thousand \\
		\bottomrule
	\end{tabular}
	\label{tab:dataset:summary}
\end{table}

\begin{figure}[h]
	\centering
	\includegraphics[scale=0.6]{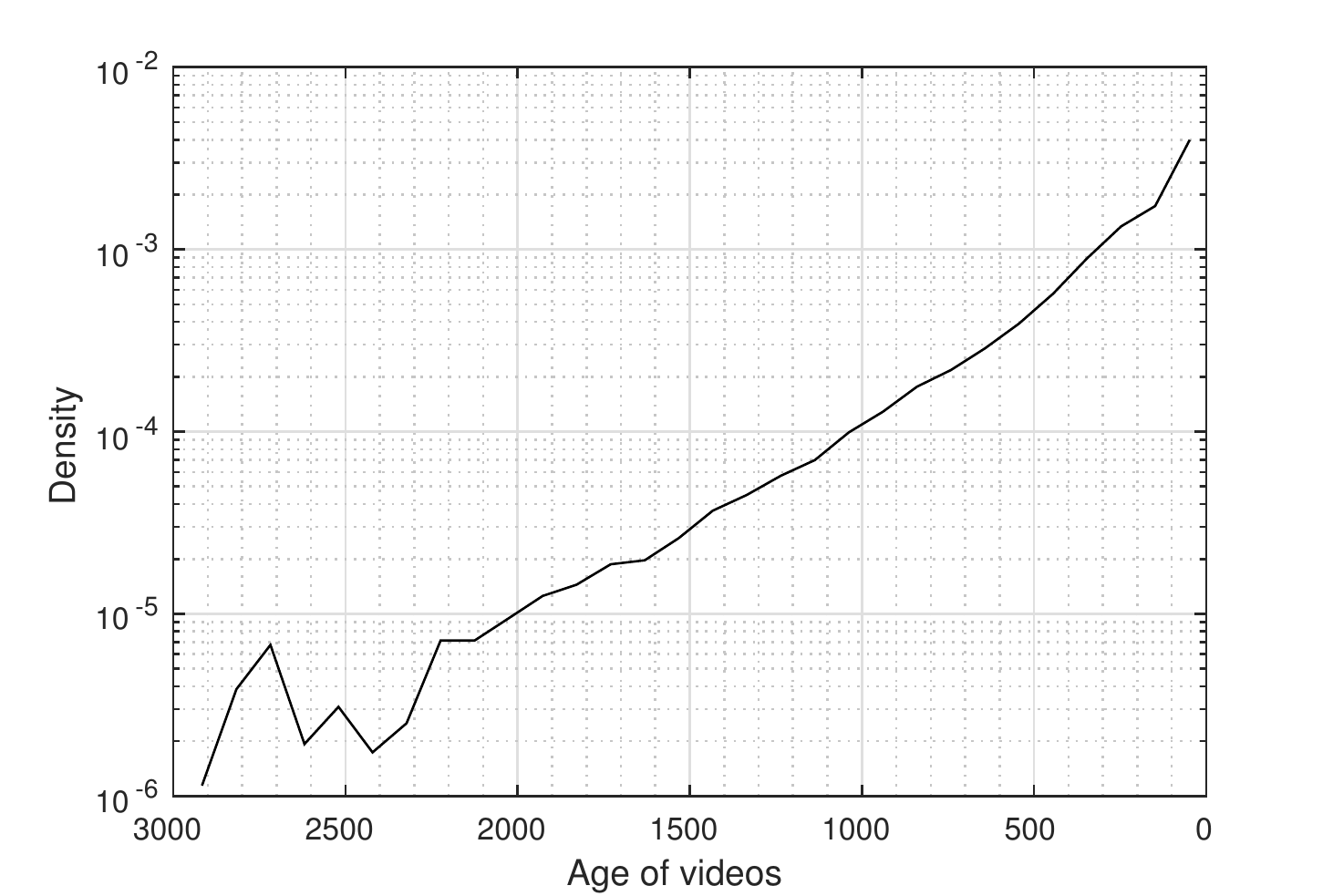}
	\caption{The fraction of videos in the dataset as a function of the age of the videos. There is a significant percentage of newer videos (videos with less age) compared to older videos. Hence, the dataset capture the exponential growth of the number of videos uploaded to YouTube. }
	\label{fig:hist:age}
\end{figure}

Table~\ref{tab:category:dataset:summary}, shows the summary of the various category of the videos present in the dataset. The dataset contains a large percentage of gaming videos.    
\begin{table}[!h]
	\centering
	\caption{YouTube dataset categories (out of $6$ million videos) }
	\begin{tabular}{c|c}
		\toprule
		Category & Fraction \\
		\midrule
		Gaming & 0.69 \\
		Entertainment & 0.07 \\
		Food & 0.07 \\
		Music & 0.035 \\
		Sports & 0.017\\
		\bottomrule
	\end{tabular}
	\label{tab:category:dataset:summary}
\end{table}
Fig.~\ref{fig:hist:age} shows the fraction of videos as a function of the age of the videos. 
There is a large fraction of videos uploaded within a year. 
Also, the dataset captures the exponential growth in the number of videos uploaded to YouTube. 
Similar to~\cite{RAEJLP14}, we define three categories of videos based on their popularity: Highly popular, popular, and unpopular. 
Table~\ref{tab:popularity:dataset:summary} gives a summary of the fraction of videos in the dataset belonging to each category. 
As can be seen from Table~\ref{tab:popularity:dataset:summary}, the majority of the videos in the dataset belong to the popular category.  
\begin{table}[!h]
	\centering
	\caption{Popularity distribution of videos in the dataset}
	\begin{tabular}{c|c}
		\toprule
		Criteria & Fraction \\
		\midrule
		Highly Popular (Total Views $> 10^4$)  &  0.12\\
		Popular ($150 < $  Total Views $< 10^4$)&  0.67\\
		Unpopular (Total Views $< 150$) &  0.21\\
		\bottomrule
	\end{tabular}
	\label{tab:popularity:dataset:summary}
\end{table}

A unique feature of the dataset is that it contains information about the ``\metalevel optimization'' for videos.  
The \metalevel optimization is a change in the title, tags or thumbnail, of an existing video in order to increase the popularity.  
BBTV markets a product that intelligently automates the \metalevel optimization. 
Table~\ref{tab:optimization:dataset:summary} gives a summary of the statistics of the various \metalevel optimization present in the dataset. 
\begin{table}[!h]
	\centering
	\caption{Optimization summary statistics}
	\begin{tabular}{c|c}
		\toprule
		Optimization & \# Videos \\
		\midrule
		Title change &  $21$ thousand \\
		Thumbnail change &  $13$ thousand \\
		Keyword change &  $21$ thousand\\
		\bottomrule
	\end{tabular}
	\label{tab:optimization:dataset:summary}
\end{table}

\section*{Acknowledgment}
The authors are grateful to Dr.\ Di Xu, Dr.\ Lino Coria and Dr.\ Mehrdad Fatourechi from BBTV for supplying the YouTube dataset described above, and also for useful discussions and reviewing a preliminary draft of this paper. 
\bibliographystyle{IEEEtran}

\end{document}